\documentclass[aps,prb,superscriptaddress,twocolumn]{revtex4-1}
\usepackage{bm}
\usepackage[fleqn]{amsmath}
\usepackage{hyperref}
\usepackage{graphicx}

\begin{document}

\title{Edge states and spin-valley edge photocurrent in transition metal dichalcogenide monolayers}

\author{V.V. Enaldiev}
\email{vova.enaldiev@gmail.com}
\affiliation{Kotel'nikov Institute of Radio-engineering and Electronics of the Russian Academy of Sciences, 11-7 Mokhovaya St, Moscow, 125009 Russia}

\date{\today}

\begin{abstract}
We develop an analytical theory for edge states in monolayers of transition metal dichalcogenides based on a general boundary condition for a two-band ${\bf kp}$-Hamiltonian in case of uncoupled valleys. Taking into account {\it edge} spin-orbit interaction we reveal that edge states, in general, have linear dispersion that is determined by three real phenomenological parameters in the boundary condition. In absence of the edge spin-orbit interaction, edge states are described by a single real parameter whose sign determines whether their spectra intersect the bulk gap or not. In the former case we show that illumination by circularly polarised light results in spin and valley polarised photocurrent along the edge. Flow direction, spin and valley polarisation of the edge photocurrent are determined by the direction of circular polarisation of the illuminated light.
\end{abstract}

\maketitle

\section{Introduction}

The optical properties of monolayer crystals of transition metal dichalcogenides (TMDs) (like MoS$_2$, MoSe$_2$, MoTe$_2$, WS$_2$, and WSe$_2$) have recently attracted a considerable interest due to possible optoelectronic applications\cite{bib:Sun,bib:Posp}. This is due to the direct band gap of the TMD monolayers whose value corresponds to the visible and infrared light frequencies\cite{bib:Falko}. The optical response of the bulk materials at absorption edge is dominated by excitons\cite{bib:Moody}. However, recent experiments on second harmonic generation at sub-band gap frequencies \cite{bib:Yin}, scanning tunneling microscopy and spectroscopy\cite{bib:Zhang,bib:Bollinger2001} as well as microwave impedance microscopy\cite{bib:Wua} in MoS$_2$-monolayers have also exhibited edge state (ES) signs.  
 
From theoretical point of view properties of ESs in atomically thin MoS$_2$ are extensively studied in frames of density functional theory\cite{bib:Boll,bib:Bollinger2001,bib:Ataca,bib:Erdogan,bib:Vojv,bib:Li}  as well as tight-binding approximation\cite{bib:Khoeini,bib:Guinea} . However description of the ESs in the TMD monolayers within ${\bf kp}$-approach allows one to describe the ESs without going into details of edge microscopic structure and to take into account effects of external fields. This enables to construct an analytic theory for the ESs in the whole class of materials in a unified way. Such a general theory relies on a boundary condition (BC) that describes the edge structure by means of several phenomenological parameters. The values of these parameters can be obtained by fitting with experimental data or other calculations based on density-functional or tight-binding approximations. Authors of Ref.[\onlinecite{bib:Korman}] derive a general boundary condition taking into account valley coupling at the edge  and neglecting by edge spin-orbit interaction which may, in general, exist (see Ref.[\onlinecite{bib:Volkov}] and references therein). Recently, ES spectra in the TMD monolayer nanoribbon\cite{bib:Segarra} and optical absorption in TMD nanoflakes involving transitions between the bulk and edge states\cite{bib:Trushin} have been studied in the ${\bf kp}$-approximation. However, these studies were restricted by some certain values of the phenomenological boundary parameters.
 
The aim of this paper is twofold. First, we develop an analytical theory for the ESs in the TMD monolayers in the ${\bf kp}$-approach taking into account {\it edge} spin-orbit interaction (ESOI) in the case of uncoupled valleys. Second, we consider optoelectronic properties of the ESs and demonstrate an emergence of spin and valley polarised edge photocurrents due to illumination of the monolayer by circularly polarised light. Origin of the effect concerns with selection rules for optical transitions from bulk states to edge states caused by circularly polarised light in the two valleys. The selection rules are essentially different from those for interband optical transitions in the bulk that give rise, for example, to the valley Hall effect\cite{bib:Mak2014}. The paper is organized as follows: in Sec.\ref{sec1} we derive a general boundary condition and resulting ES spectra for two-band continuum model, Sec.\ref{Sec2} is devoted to derivation of the edge photocurrent.

\section{Edge states in two-band {\bf kp}-approximation}\label{sec1}
%{\it The Model and Edge State Spectra.} 
In the TMD monolayers conduction and valence band edges are located in $K$ and $K'$ valleys of the honeycomb lattice Brillouin zone. Within a two-band ${\bf kp}$-approach, dynamics of electrons in the $K$ $(K')$ valley is described by the Hamiltonian\cite{bib:Falko,bib:Xiao}
\begin{equation}\label{TMD_Hamiltonian}
H_{\tau} = \left( 
\begin{array}{cccc}
m + \tau\Delta_c & vp_{-,\tau} & 0 & 0 \\
vp_{+,\tau} & -m + \tau\Delta_v & 0 & 0 \\
0 & 0 &  m - \tau\Delta_c & vp_{-,\tau} \\
0 & 0 & vp_{+,\tau} & -m - \tau\Delta_v 
\end{array}
\right)
\end{equation}
where $2m$ is the band gap without spin splitting, $2\Delta_{c,v}$ is the value of spin splitting in the conduction and valence band correspondingly, the index $\tau=+1(-1)$ denotes the $K$ $(K')$ valley, $p_{\pm,\tau} = \tau p_x \pm ip_y$ ($p_x, p_y$ are components of in-plane momentum), $v$ is the velocity matrix element between the band extrema. The Hamiltonian $H_{\tau}$ (\ref{TMD_Hamiltonian}) possesses diagonal form in the spin subspace, with the upper-left (lower-right) block acting on two-component wave functions of spin up (down) states. To describe  edge of the TMD monolayer one should supplement the Hamiltonian with a BC for envelope wave functions. In present work we consider a translation invariant edge for which projections of the valley centers onto the edge direction are well distant from each other (like at zigzag or reconstructed zigzag edges). Therefore, we will neglect by the valley coupling at the edge. 

\begin{figure}
\includegraphics[width=8.5cm,height=5cm]{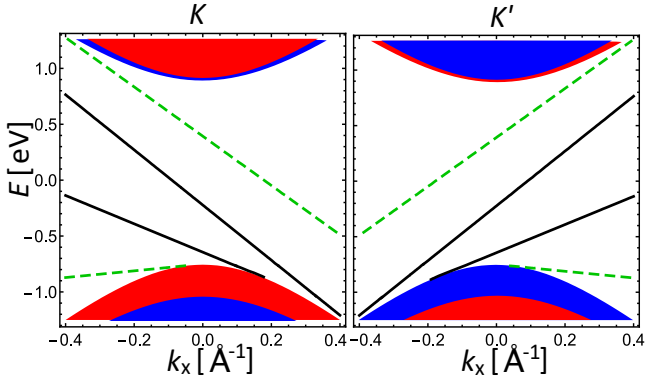}
\caption{\label{Fig1} Solid black and dashed green lines represent typical spectra of ESs in the $K$ (left) and $K'$ valleys (right) derived from Eq. (\ref{gen_Disp_eq}) for ESOI described by the following phenomenological parameters: Dashed green lines respond to $\xi=-2.9$, $\eta=0.5$, $\nu=0.6$; solid black lines respond to $\xi=-2.9$, $\eta=1.5$, $\nu=0.6$. Red and blue shaded regions shows projections of spin up and spin down bulk bands correspondingly. Bulk parameters are as follows: $2m=1.8$ eV, $\Delta_c=10$ meV, $\Delta_v=143$ meV, $v=2.5$ eV$\cdot$A.}
\end{figure}

However, even at the translation invariant edge, additional ESOI can mix the spins. Let us derive the most general BC that describes this entanglement in our model. Since $H_{\tau}$ (\ref{TMD_Hamiltonian}) is of the first order in momentum, a general BC has a form of a linear combination of two-component wave functions $\Psi^{\uparrow(\downarrow)}_{\tau} = (\psi^{\uparrow(\downarrow)}_{c,\tau},\psi^{\uparrow(\downarrow)}_{v,\tau})^T$ belonging to spin up (down) states:
\begin{equation}\label{gen_BC}
\left[\Psi^{\uparrow}_{\tau} - M_{\tau}\Psi^{\downarrow}_{\tau} \right]_{\rm edge} = 0
\end{equation}
here $M_\tau$ is the second order square matrix consisting of phenomenological parameters that characterize the edge structure. Explicit form of the matrix $M_{\tau}$ is determined by the Hermiticity of the Hamiltonian $H_{\tau}$ in a confined region:
\begin{equation}\label{Hermiticity}
\left[M^{+}_{\tau}{\bf n}{\bm \sigma}_{\tau} M_{\tau} + {\bf n}{\bm \sigma}_{\tau}\right]_{\rm edge} = 0 
\end{equation}
where $\bm{\sigma}_{\tau} = (\tau\sigma_x,\sigma_y)$ is vector of Pauli matrix, ${\bf n}=(n_x,n_y)$ is a outer unit  normal to the edge. Eq.(\ref{Hermiticity}) can also be obtained from requirement of vanishing normal component of probability current at the edge\cite{bib:Akhmerov}. Time reversal symmetry relates the matrices in BC (\ref{gen_BC}) from the two valleys: $M_{\tau}^{-1} = M_{-\tau}^*$. These conditions lead us to the four-parametric form for matrix $M_{\tau}$:
\begin{equation}\label{M_matrix}
\begin{split}
M_{\tau} =
ie^{i\chi}\left[\sigma_0\sinh\xi +  \tau\sigma_z\cosh\xi\cosh\eta\cos\nu -\right. \qquad\qquad\qquad\\
\left. i\tau {\bf n}\bm{\sigma}_{\tau}\cosh\xi\left(\sinh\eta + \sigma_z\cosh\eta\sin\nu\right) \right]\quad\,
\end{split}
\end{equation}
where $\sigma_0,\sigma_z$ are identity and the third Pauli matrices respectively, $\chi,\xi,\eta,\nu$ (\mbox{$0\leq \chi,\nu<2\pi$}, \mbox{$-\infty<\xi,\eta<+\infty$}) are real phenomenological parameters characterizing edge properties. 
\begin{figure}
	\includegraphics[width=8.5cm,height=5cm]{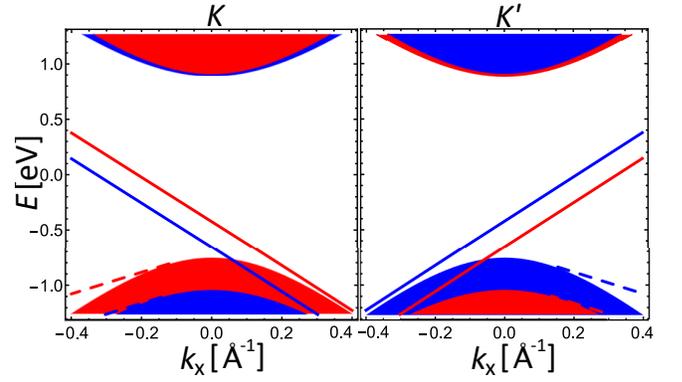}
	\caption{\label{Fig2} Red (blue) lines show spectra of spin up (down) ESs (\ref{ES_spectrum}) in the $K$ (left) and $K'$ valleys (right) in absence of ESOI (i.e. $a_{+1,\tau}=a_{-1,\tau}$). Solid lines respond to $a_{\pm 1,\tau}=0.5$; dashed lines respond to $a_{\pm 1,\tau}=-0.2$. Red and blue shaded regions show projections of spin up and spin down bulk bands correspondingly. Bulk parameters are the same as on Fig. \ref{Fig1}.}
\end{figure}
As matrix $M_{\tau}$ depends on the parameter $\chi$ through a common phase factor $e^{i\chi}$ that we assume is constant as a function of coordinate along translation invariant edge, we eliminate it by means of a unitary transformation $(\widetilde{\Psi}^{\uparrow}_{\tau}, \widetilde{\Psi}^{\downarrow}_{\tau})^{\rm T} = U (\Psi^{\uparrow}_{\tau}, \Psi^{\downarrow}_{\tau})^{\rm T}$ where $U={\rm diag}\left\{\sigma_0,e^{-i\chi}\sigma_0\right\}$. Thus, only three boundary parameters $\xi,\eta,\nu$ characterize translation invariant edge in the TMD monolayer in absence of intervalley interaction. Their physical meaning may be obtained by considering limiting cases of the BC (\ref{gen_BC}). In the limit \mbox{$\xi\to+\infty$}, two-component wave functions with opposite spins become decoupled and satisfy BC (for derivation, see Appendix \ref{App_BC}):
\begin{equation}\label{BC_one_param}
\left[\psi_{c,\tau}^{\uparrow (\downarrow)} - i a_{s,\tau} e^{-i\tau\varphi} \psi_{v,\tau}^{\uparrow (\downarrow)}\right]_{{\rm edge}}=0
\end{equation}
where $a_{s,\tau} = s\tau (1 + s\tau \cosh\eta\cos\nu)/(\sinh\eta+\cosh\eta\sin\nu)$, $\varphi$ is an angle characterizing the unit normal \mbox{${\bf n}=(\cos\varphi,\sin\varphi)$}, $s=+1, (-1)$ for spin up (down) states. However, disentanglement of the wave functions belonging opposite spin projections in BC (\ref{BC_one_param}) does not mean vanishing ESOI, as $a_{1,\tau}\neq a_{-1,\tau}$ in general case. It is known\cite{bib:Zag} that BC (\ref{BC_one_param}) is equivalent to insertion of diagonal in spin subspace potential in the Hamiltonian (\ref{TMD_Hamiltonian}), which is a combination of electrostatic ($\propto \sigma_0$) and pseudo-electrostatic ($\propto \sigma_z$) potentials $V^{\uparrow(\downarrow)}_{\rm edge}=[V({\bf r})/2]\left[\sigma_0\left(1- a_{s,\tau}^2\right) + \sigma_z\left( 1 + a_{s,\tau}^2\right) \right]$, where $V({\bf r})$ tends to infinity outside the 2D material and is zero inside of it. The sign of $a_{s,\tau}$ is determined by the sign of $V({\bf r})$ outside of TMD monolayer. In addition, at $\eta\to+\infty$ in BC (\ref{BC_one_param}), values of the boundary parameters $a_{\pm 1,\tau}$ coincide and equal to $a=\cos\nu/(1+\sin\nu)$, which is determined only by the parameter $\nu$. Therefore, we can conclude that $\xi$ and $\eta$ describe different types of ESOI, like Rashba and Dresselhaus {\it interface} spin-orbit parameters for Schrodinger's electrons in GaAs/Al$_x$Ga$_{1-x}$As quantum wells\cite{bib:Deviz}, but $\nu$ is responsible for coupling of bands with the same spin.

Now we are able to calculate spectra of ESs with the general BC (\ref{gen_BC}),(\ref{M_matrix}). Suppose that the 2D crystal fills a half-plane $y>0$. Then, ES wave function is of the form:
\begin{equation}\label{ES_wave_function}
\Psi^{\uparrow(\downarrow)}_{\tau}=C^{\uparrow(\downarrow)}_{\tau}
\left(
\begin{array}{c}
1 \\
\frac{\hbar v\left(\tau k_x - \kappa^{\uparrow(\downarrow)}_{\tau}\right)}{\varepsilon + m - s\tau\Delta_v }
\end{array}
\right)e^{-\kappa^{\uparrow(\downarrow)}_{\tau}y + ik_xx}
\end{equation}
where $C^{\uparrow(\downarrow)}_{\tau}$ is a normalization constant,
\mbox{$\kappa^{\uparrow(\downarrow)}_{\tau}=\left\{k_x^2 - (\varepsilon - m + s\tau\Delta_c)(\varepsilon + m - s\tau\Delta_v)/(\hbar v)^2\right\}^{1/2}$} is a decay length of ESs. Substituting wave function (\ref{ES_wave_function}) in the BC (\ref{gen_BC}) we obtain a general dispersion equation for ESs:
\begin{equation}\label{gen_Disp_eq}
\begin{split}
\left[1 + a_{+1,\tau}\frac{\hbar v\left(\tau k_x - \kappa^{\uparrow}_{\tau}\right)}{\varepsilon + m - \tau\Delta_v} \right]
\left[ 1 +  a_{-1,\tau}\frac{\hbar v\left(\tau k_x - \kappa^{\downarrow}_{\tau}\right)}{\varepsilon + m + \tau\Delta_v}\right] + \qquad  \\
\frac{\tau v\left(\tanh\xi - 1\right)}{\sinh\eta+\cosh\eta\sin\nu}\left[\frac{\hbar v\left(\tau k_x - \kappa^{\uparrow}_{\tau}\right)}{\varepsilon + m - \tau\Delta_v} -  %\right.  \\ 
%\left. 
\frac{\hbar v\left(\tau k_x - \kappa^{\downarrow}_{\tau}\right)}{\varepsilon + m + \tau\Delta_v}\right] = 0. 
\end{split}
\end{equation}

On Fig.\ref{Fig1} we reveal typical dispersion of ESs in the  $K$ and $K'$ valleys given by the previous equation. In general case, ESs possess linear dispersion and exist for those longitudinal momenta when their energies are not overlapped with projections of bulk bands. The second term in the left-hand side of Eq.(\ref{gen_Disp_eq}) is responsible for coupling of spins due to ESOI as it goes to zero at $\xi\to+\infty$, which is the discussed above limit of spin disentanglement. In this limit, ES spectra are determined by its own parameter $a_{s,\tau}$ for each spin projection:
\begin{equation}\label{ES_spectrum}
\varepsilon^{\uparrow(\downarrow)}_{e,\tau}(p_x) = -\tau \widetilde{v}_{s,\tau}p_x + \varepsilon_{s,\tau} ,
\end{equation}
where \mbox{$\widetilde{v}_{s,\tau}=2a_{s,\tau}v/(1+a_{s,\tau}^2)$} is an effective speed of ESs, \mbox{$\varepsilon_{s,\tau}=\widetilde{m}_{s,\tau}(a_{s,\tau}^2-1)/(a_{s,\tau}^2+1) + s\tau(\Delta_c+\Delta_v)/2$} is energy of ESs at center of the corresponding valley, and \mbox{$\widetilde{m}_{s,\tau} = m - s\tau\left(\Delta_v - \Delta_c\right)/2$}. Valley index $\tau$ determines chirality of ESs (see Fig. \ref{Fig2}) which exist for those wave vectors $k_x$ while their decay length is positive:
\begin{equation}\label{kappa}
\kappa^{\uparrow(\downarrow)}_{\tau} = \frac{\tau k_x}{\widetilde{m}_{s,\tau}}\left(\varepsilon_{s,\tau}- s\tau\frac{\Delta_v+\Delta_c}{2}\right) + \frac{\widetilde{m}_{s,\tau}\widetilde{v}_{s,\tau}}{\hbar v^2}>0.
\end{equation}
We note that condition (\ref{kappa}) allows spin-polarised ESs to exist even when their spectra are overlapped with projection of bulk band characterized by opposite spin. In particular, inequality (\ref{kappa}) specifies that at $a_{s,\tau}>0$ ES spectra intersect the bulk gap, but at $a_{s,\tau}<0$ they are out of the gap (see Fig. \ref{Fig2}). Wave functions of ESs with spectrum (\ref{ES_spectrum}) read as follows:
\begin{equation}\label{ES_function}
\Psi^{\uparrow(\downarrow)}_e = C_{\tau}^{\uparrow(\downarrow)}
\left (
\begin{array}{c}
1 \\
-\frac{\tau}{a_{s,\tau}} 
\end{array}
\right )e^{-\kappa^{\uparrow(\downarrow)}_{\tau} y + ik_xx},
\end{equation}
where $C_{\tau}^{\uparrow(\downarrow)}=\left[2a_{s,\tau}^2\kappa^{\uparrow(\downarrow)}_{\tau}/L_x(1+a_{s,\tau}^2)\right]^{1/2}$ is the normalization factor. 

Here we point out that in absence of ESOI (\mbox{$\xi,\eta\to+\infty$}), ES spectra for two spins are parallel to each other with constant energy difference for every momenta: \mbox{$\varepsilon_{e,\tau}^{\uparrow} - \varepsilon_{e,\tau}^{\downarrow} =  2\tau\left( \Delta_v + a^2\Delta_c\right)/(1+a^2)$}.

\section{Spin-valley edge photocurrents}\label{Sec2}

In this section we reveal that illumination of semi-infinite 2D TMD crystal by circularly polarised light induces spin and valley polarised edge photocurrents. As we mentioned in Introduction the effect is due to difference in transition probabilities between bulk and edge bands in the two valleys caused by the light. 

\begin{figure}
\includegraphics[width=9cm]{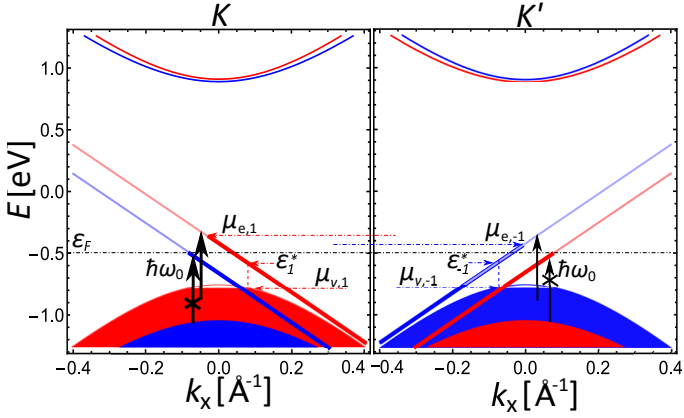}
\caption{\label{Fig3} Schematic picture of quasi Fermi-levels in edge ($\mu_{e,\pm 1}$) and valence ($\mu_{v,\pm 1}$) bands under illumination of clockwise circular polarised light with frequency $\omega_0$ in case of spin polarised ESs described by the formula (\ref{ES_spectrum}) with $a_{+,\tau}=a_{-,\tau}>0$. $\varepsilon_F$ is equilibrium Fermi-energy. $\varepsilon_{\pm 1}^{*}$ is the minimal energy of ESs that radiatively recombine with holes in valence band. }
\end{figure}

For definiteness and simplicity we consider absence of spin-orbit interaction at the edge. Therefore ESs are described by a single parameter $a$ for both spins and valleys (we suppressed spin and valley indexes). However, obtained results are also valid when the two spin-polarised ES branches are described by different parameters $a_{+1,\tau}\neq a_{-1,\tau}$ (\ref{ES_spectrum}), while one can populate only one of them in each valley.  We will regard that $a\sim 1$, so that ES spectra intersect the gap as shown on Fig.\ref{Fig3}. This situation  agrees with tight-binding calculations of ES spectra at zigzag edge of MoS$_2$\cite{bib:Guinea}. Further throughout this section we suppress the valley index $\tau$ everywhere except where it is needed.

Now we turn to the bulk states. Below we are only interested in valence band states with energies around band extremum. In this limit ($vp\ll 2\widetilde{m}$) spectra of the spin up (down) states in valence band of the $K$ $(K')$ valley are expressed as follows:
\begin{equation}\label{valence_energies}
\varepsilon_{v} = \varepsilon_t - \frac{(vp)^2}{2\widetilde{m}},
\end{equation}
where $\varepsilon_t=-m + \Delta_v $ is energy of the upper valence band top in each valley, $p=\sqrt{p_x^2 + p_y^2} $ is 2D momentum modulus. Under assumption about type of the edge mentioned above, bulk states should satisfy the BC (\ref{BC_one_param}) with the same parameter $a$ for both spins. This BC is satisfied by superposition of incident plane wave and plane wave scattered off the edge with a common longitudinal momentum $p_x$:
\begin{equation}\label{bulk_solution}
\Psi_{p_x,\varepsilon_v} = \frac{1}{\sqrt{2}}
\left[
\psi_{p_x,-p_y} + R_{\varepsilon_v,k_x}\psi_{p_x,p_y}
\right]
\end{equation}
where $\psi_{p_x,\pm p_y}$ are plane wave solutions of the Hamiltonian (\ref{TMD_Hamiltonian}) in the limit \mbox{$vp\ll 2\widetilde{m}$}, $R_{\varepsilon_v,k_x}$ is a reflection coefficient that is determined by the BC. Plane wave states around extremum of valence band are of the form (see Appendix \ref{AppPlaneWaves}):
\begin{equation}\label{plane_wave}
\psi_{p_x,p_y}=\frac{1}{\sqrt{L_xL_y}}
\left(
\begin{array}{c}
-\dfrac{vp}{2\widetilde{m}} \\
\\
e^{i\vartheta_p}
\end{array}
\right)e^{i{\bf k}{\bf r}},
\end{equation}
where $L_{x}L_{y}$ is the system area, \mbox{$e^{i\vartheta_p}=p_{+,\tau}/p$}, \mbox{${\bf k}=\left(p_x,p_y\right)/\hbar$} is 2D wave vector. In fact in semi-infinite sample $p_y$ is not a good quantum number and should be treated as a function of energy and $p_x$ from Eq. (\ref{valence_energies}): \mbox{$p_y(\varepsilon,p_x)=[\sqrt{2\widetilde{m}}/v]\sqrt{\varepsilon_t - \varepsilon_v-(vp_x)^2/2\widetilde{m}}$}. For valence band states with energies around band extremum reflection coefficient can be approximated as follows $R_{\varepsilon_v,k_x}\approx -e^{-i2\theta_{p}}$.

We suppose that the semi-infinite 2D crystal is illuminated in negative direction of $z$-axis by a clockwise polarised light with frequency $\omega_0$ (in case of counter clockwise polarisation one should exchange $\tau\to-\tau$ in final formulae (\ref{equation_mu_el}),(\ref{mu_e})). Due to spin splitting of the valence bands (order of $0.1$ eV) one can tune frequency $\omega_0$ of illuminated light and Fermi-energy $\varepsilon_F$ in the monolayer, to induce electrical dipole transitions only from the upper valence band in the $K$ ($K'$) valley to ESs with the corresponding spin (see Fig.\ref{Fig3}). Our aim is to calculate induced edge photocurrent owing to these transitions. To this end we derive kinetic equation for distribution function in frames of Keldysh formalism\cite{bib:Arseev} (see Appendix \ref{AppA}):
\begin{equation}\label{kin_eq}
\begin{split}
\dfrac{\partial f_e(\varepsilon)}{\partial t} =  W^{ind}_{\varepsilon-\hbar\omega_0,\varepsilon}\left[f_v(\varepsilon - \hbar\omega_0) - f_e(\varepsilon) \right] - \qquad\qquad\qquad\\
f_e(\varepsilon)\int_0^{+\infty}  W^{sp}_{\varepsilon-\hbar\omega,\varepsilon}\left[1-f_v(\varepsilon - \hbar\omega)  \right]d\omega - 
\dfrac{f_e(\varepsilon)-f_{eq}(\varepsilon)}{\tau_R},
\end{split}
\end{equation}  
where $f_{e,v}(\varepsilon)$ is Fermi-Dirac distribution function for states in edge/valence band respectively, $f_{eq}(\varepsilon)$ is equilibrium Fermi-Dirac function, $W^{ind}_{\varepsilon-\hbar\omega_0,\varepsilon}$ is a rate of induced transitions between ES, characterized by longitudinal momentum $p_e$ and energy $\varepsilon$, and the valence band state with the same longitudinal momentum and energy $\varepsilon-\hbar\omega_0$, $W^{sp}_{\varepsilon-\hbar\omega,\varepsilon}d\omega$ is a rate of spontaneous transitions caused by interaction with ground state of electromagnetic field, $\tau_R$ is a phenomenological relaxation time that describes other relaxation processes between edge and valence band states caused by, for example, phonon or electron-electron scattering (below we discuss range of the relaxation time). The rate of the induced transitions reads as follows
\begin{equation}\label{prob_ind}
W^{ind}_{\varepsilon-\hbar\omega_0,\varepsilon} = \frac{2\pi}{\hbar}\frac{I\Omega}{c\hbar \omega_0}\left|V^{(+,\tau)}_{\varepsilon-\hbar\omega_0,\varepsilon}({\bf q}_0)\right|^2\rho_{v,e}\left(\varepsilon - \hbar\omega_0\right)
\end{equation}
%$I=cn_r^2\omega_0^2A_0^2/4\pi$
where $I$ is intensity of the incident light, $c$ is the speed of light, $\Omega$ is a volume for quantization of electromagnetic field, $V^{(+,\tau)}_{\varepsilon-\hbar\omega_0,\varepsilon}({\bf q}_0)$ is the matrix element of interaction between valence and edge band states (\ref{App_field_matrix_element}), density of the valence band states with definite longitudinal momentum $p_e$ near valence band extrema is expressed by the formula:
\begin{equation}\label{DOS_v}
\rho_{v,e}(\varepsilon)=\sum_{p_y}\delta\left(\varepsilon - \varepsilon_{p_e,p_y}\right) = \dfrac{L_y\sqrt{2\widetilde{m}}}{2\pi\hbar v}\dfrac{\Theta\left(\varepsilon_{t} - \frac{v^2p_e^2}{2\widetilde{m}} - \varepsilon\right)}{\sqrt{\varepsilon_{t} - \frac{v^2p_e^2}{2\widetilde{m}} - \varepsilon}},
\end{equation}
where $\Theta(...)$ is the Heaviside step function. The probability of spontaneous transitions is expressed as follows:
\begin{equation}\label{prob_sp}
W^{sp}_{\varepsilon-\hbar\omega,\varepsilon} = \frac{\Omega}{(2\pi)^2\hbar c^3}\sum_{\lambda=\pm}\int do_q\left|V^{(\lambda,\tau)}_{\varepsilon-\hbar\omega,\varepsilon}({\bf q})\right|^2 \rho_{v,e}\left(\varepsilon - \hbar\omega\right),
\end{equation}
here integration goes over solid angle of light wave vector ${\bf q}$, summation runs over clockwise ($\lambda=+$) and counter clockwise ($\lambda=-$) polarisation of the electromagnetic field.

It is known that intraband energy relaxation processes have the shortest times in MoS$_2$ monolayers\cite{bib:Shi,bib:Wang} (order of picoseconds). As electron-electron scattering is very efficient in one dimension\cite{bib:Hirtschulz} and also due to possibility of relaxation in edge band via scattering by bulk phonons, we suppose that relaxation time within the edge band is of the same order. This allows us to solve the kinetic equation (\ref{kin_eq}) in quasi-equilibrium approximation, which is justified when intra-band relaxation times is shorter than edge-bulk energy relaxation times. Therefore, we look for distribution function of the edge (valence) states in the form of Fermi-Dirac function with its own quasi Fermi-level $\mu_{e,\tau}$ ($\mu_{v,\tau}$). We also suppose that additional edge-to-valence band energy relaxation processes, characterized by the relaxation time $\tau_R$ order of nanoseconds. This is mainly due to suppression of phase-space for these processes caused by great difference in density of bulk and edge states. This allows edge-to-valence scattering only with certain longitudinal momentum mismatch (as valence band quasi Fermi-level locates around the top of valence band).

\begin{figure}
\includegraphics{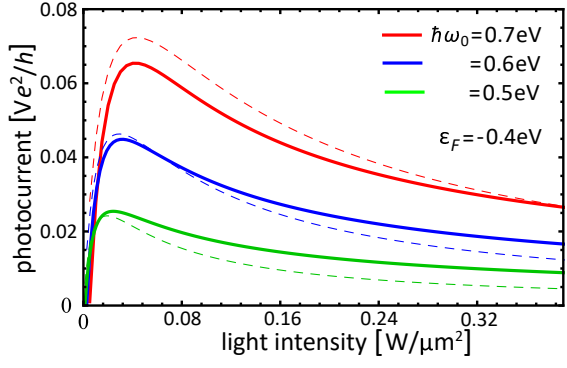}%[width=9cm]
\caption{\label{Fig4} Dependence of photocurrent (\ref{photocurrent}) on intensity of illuminating light at three different frequencies $\hbar\omega_0=0.7$ eV (red), $\hbar\omega_0=0.6$ eV (blue) and $\hbar\omega_0=0.5$ eV (green). Solid lines represent solution for quasi Fermi-levels derived from Eq.(\ref{App_relation}). Dashed lines correspond to the approximated solution (\ref{mu_e}). Equilibrium Fermi-energy is $\varepsilon_F=-0.4$ eV. Bulk parameters are similar to Fig.\ref{Fig1}, boundary parameter $a=0.8$. Phenomenological relaxation time $\tau_R=1$ ns, length characterizing the TMD monolayer in perpendicular to the edge derection is $L_y=1$ $\mu$m.}
\end{figure}

We are interested in stationary solution that equates the right-hand side of Eq.(\ref{kin_eq}) to zero. After integration the kinetic equation over energy in the limit of zero temperature we arrive to the first implicit relation for $\mu_{e,\tau}$ and $\mu_{v,\tau}$ (\ref{App_relation}). Another relation between quasi Fermi-levels is imposed by a particle conservation rule (i.e. the number of holes in the valence band equals the number of photo-excited electrons in the edge band): 
\begin{equation}\label{eh-conservation}
\left(\varepsilon_t-\mu_{v,\tau}\right)\rho_v = \left(\mu_{e,\tau} - \varepsilon_F\right)\rho_e
\end{equation}
where $\rho_v=L_yL_x\widetilde{m}/2\pi\left(\hbar v\right)^2$ is density of states in valence band, $\rho_e=L_x/2\pi\hbar |\widetilde{v}| $ is density of edge states. By virtue of ratio between the densities $\rho_e/\rho_v\propto \hbar v/\widetilde{m}L_y\ll 1$ one has inequality $|\varepsilon_{v}-\mu_{v,\tau}|\ll |\mu_{e,\tau}-\varepsilon_F|$, which leads us to a simpler equation for $\mu_{e,\tau}$ (in comparison with Eq.(\ref{App_relation})):
\begin{equation}\label{equation_mu_el}
\begin{split}
\dfrac{I}{I_{0}}\left[\delta_{\tau,1}+\delta_{\tau,-1}\left(\dfrac{\hbar v \kappa_F}{2\widetilde{m}a}\right)^2\right]
G\left(\mu_{e,\tau}-\hbar\omega_0\right) - 
\dfrac{\mu_{e,\tau}-\varepsilon_F}{\hbar\omega_0^2\widetilde{\tau}_R} = 0, \qquad\qquad\qquad\qquad\qquad\qquad\qquad\qquad\qquad
\end{split}
\end{equation}
here $I_0=\hbar\omega_0^4/c^2$, \mbox{$\widetilde{\tau}_R=\tau_R\left[(v/c)^2(e^2/\hbar c)4\pi a^2/(1+a^2)n_r^2\right]$} ($n_r$ is refractive index of environment), and
\begin{equation}
G(\varepsilon) = \arctan\left(\dfrac{k_{y0}(\varepsilon)}{\kappa_F}\right) - \dfrac{\kappa_F k_{y0}(\varepsilon)}{\kappa_F^2 + k_{y0}^2(\varepsilon)},\nonumber
\end{equation}
where \mbox{$k_{y0}(\mu_{e,\tau}-\hbar\omega_0)=p_y(\mu_{e,\tau}-\hbar\omega_0,p_x=0)/\hbar$}.
Deriving Eq.(\ref{equation_mu_el}) we neglected the dependence of decay length of ESs on the energy (i.e. \mbox{$\kappa(\varepsilon)\approx\kappa(\varepsilon_F)\equiv\kappa_F$}). We also disregarded dependence of momentum  component $p_y(\varepsilon,p_e)$ on $p_e$, since we consider the valence band states around the band extremum, i.e. with longitudinal momenta \mbox{$vp_e\ll 2\widetilde{m}$}. It is the expression in the square brackets of Eq.(\ref{equation_mu_el}) that describes distinction in probabilities of transitions from valence band states (\ref{bulk_solution}) to ESs (\ref{ES_wave_function}) in the $K$ and $K'$ valleys, which significantly differs from selection rules for interband transitions in the bulk\cite{bib:Xiao}. This results in uncompensated edge photocurrent with a specific spin in a particular valley. In the limit $|\mu_{e,\tau}-\varepsilon_F|/\varepsilon_F \ll 1$ we obtain explicit expression for electron quasi Fermi-level in the edge band:
\begin{equation}\label{mu_e}
\begin{split}
\mu_{e,\tau} = \varepsilon_F + 
\dfrac{\frac{I}{I_0}\hbar\omega_0^2\widetilde{\tau}_R G\left(\varepsilon_F-\omega_0\right)\left[\delta_{\tau,1}+\delta_{\tau,-1}\left(\frac{\hbar v \kappa_F}{2\widetilde{m}a}\right)^2\right]}{1 - \frac{I}{I_0}\hbar\omega_0^2\widetilde{\tau}_R\frac{\partial G(\varepsilon_F-\omega_0)}{\partial \varepsilon}\left[\delta_{\tau,1}+\delta_{\tau,-1}\left(\frac{\hbar v \kappa_F}{2\widetilde{m}a}\right)^2\right]}.
\end{split}
\end{equation}
With known quasi Fermi-levels of the ESs in the two valleys, we use a standard formula for one-dimensional current along the sample edge: 
\begin{equation}\label{photocurrent}
j = \frac{e}{h}\left(\mu_{e,1} - \mu_{e,-1}\right).
\end{equation}
Dependence of the photocurrent on the light intensity is plotted on Fig.\ref{Fig4}. Solid lines show photocurrent obtained by solving Eq.(\ref{App_relation}) for electron quasi Fermi-levels, dashed lines represent approximated solution (\ref{mu_e}). At small intensities, the current (\ref{photocurrent}) is a linear function of the intensity with a tilt that are determined by the edge-valence band transition probability difference in the $K$ and $K'$ valleys. However, in the limit of high intensities the current goes to zero as edge bands in both valleys tend to equal population. It gives rise to a maximum photocurrent in middle region of intensities. Here we notice that light intensity as high as hundreds mW/$\mu$m$^2$ in continous wave regime for near-infrared range of wavelengths can be realized by means of fiber lasers\cite{bib:Ter}. Using  solutions (\ref{mu_e}) one can obtain an approximated expression for maximal current in Fig.\ref{Fig4}:
\begin{equation}\label{maximal_current}
j_{{\rm max}} = \frac{e}{h}
\left(\varepsilon_t - \varepsilon_F + \omega_0\right) \dfrac{1 - \left(\frac{\hbar v \kappa_F}{2\widetilde{m}a}\right)^2}{1 + \left(\frac{\hbar v \kappa_F}{2\widetilde{m}a}\right)^2},
\end{equation} 
which is valid in the limit $k_{y0}(\varepsilon_F-\omega_0)/\kappa_F\ll 1$. Therefore, the greater edge-valence band transition probability difference in the two valleys, the higher maximal value of spin-valley polarised current along the edge. Absolute values of the maximal current can attain several microamperes. In case of clockwise polarization of light, uncompensated photocurrent flows in the $K$ valley in negative direction along the edge (see Fig.\ref{Fig3}). As it was mentioned above for counter-clockwise polarisation one should exchange $\tau\to-\tau$ in (\ref{mu_e}), which leads to uncompensated photocurrent in the $K'$ valley but in positive direction along the edge. Thus, direction of light polarisation controls not only spin and valley polarisation of the uncompensated edge photocurrent but also its direction of flowing.

\section{Conclusion}

We developed a theory of ESs in monolayers of TMD crystals which takes into account ESOI within intra-valley approximation. The theory relies on a general BC comprising three real phenomenological parameters $\xi,\eta,\nu$ that characterize microscopic structure of the edge. We revealed that $\xi$ is responsible for spin coupling, $\eta$ accounts for inequivalence of the edge structure for opposite spin projections in case of decoupled spins, and $\nu$ describes interband interaction of states with the same spin. In general case, ESs have linear spectra determined by the all three  parameters. However, in the case of decoupled spins ($\xi\to +\infty$), ESs spectra become chiral in valley index, and are described by a single parameter (which is function of $\eta$ and $\nu$) for each spin projection. Sign of the latter parameter determines whether ES spectra are in the bulk gap or outside of it. 

We also considered optical pumping from valence band states to ESs in absence of the ESOI and demonstrate possibility for generation of spin and valley polarised edge photocurrents. We revealed that direction, valley and spin polarisation of the photocurrent are determined by direction of circular polarisation of the light. Maximal values of the photocurrent are order of several microamperes.

\acknowledgements
This work was supported by the Russian Foundation for Basic Research (Project No. 16-32-00655).

\appendix

\section{Derivation of spin-polarised BC (\ref{BC_one_param})}\label{App_BC}

In the limit $\xi\to +\infty$ the general BC (\ref{gen_BC}) with matrix $M_{\tau}$ (\ref{M_matrix}) is reduced to the following one:
\begin{equation}\label{AppBC_1}
\begin{split}
\left[\frac{-ie^{-\xi-i\chi}}{1+\tau\cosh\eta\cos\nu}
\left(
\begin{array}{cc}
1 & 0 \\
0 & ia_{1,\tau}e^{-i\tau\varphi}
\end{array}
\right)\Psi_{\tau}^{\uparrow} + \right. \qquad\qquad\\
\left. \left(
\begin{array}{cc}
1 & -ia_{-1,\tau}e^{-i\tau\varphi} \\
1 & -ia_{-1,\tau}e^{-i\tau\varphi}
\end{array}
\right)\Psi_{\tau}^{\downarrow} \right]_{\rm edge} = 0,
\end{split}
\end{equation}
where $a_{\pm 1,\tau} = \pm\tau (1 \pm \tau\cosh\eta\cos\nu)/(\sinh\eta + \cosh\eta\sin\nu)$  and we used an identity:
\begin{equation}
\frac{\sinh\eta - \cosh\eta\sin\nu}{1 + \tau\cosh\eta\cos\nu}=-\frac{1 - \tau\cosh\eta\cos\nu}{\sinh\eta + \cosh\eta\sin\nu}.\nonumber
\end{equation}
In the limit under consideration, wave functions accounting for opposite spin projections are decoupled in the BC, as the coefficient under $\Psi_{\tau}^{\uparrow}$ in Eq.(\ref{AppBC_1}) is exponentially small. This leads us to a BC for spin-down states:
\begin{equation}
\left[\psi^{\downarrow}_{c}-ia_{-1,\tau}e^{-i\tau\varphi}\psi^{\downarrow}_{v}\right]_{\rm edge} = 0.
\end{equation} 
where $\Psi^{\downarrow}_{\tau}=(\psi^{\downarrow}_{c},\psi^{\downarrow}_{v})$. Subtracting the second raw of the BC (\ref{AppBC_1}) from the first one, we arrive for a BC for spin up states:
\begin{equation}
\left[\psi^{\uparrow}_{c}-ia_{1,\tau}e^{-i\tau\varphi}\psi^{\uparrow}_{v}\right]_{\rm edge} = 0.
\end{equation} 
here $\Psi^{\uparrow}_{\tau}=(\psi^{\uparrow}_{c},\psi^{\uparrow}_{v})$.

\section{Approximate plane wave solutions of the Hamiltonian (\ref{TMD_Hamiltonian}) around bulk band extrema}\label{AppPlaneWaves}

In this section we find plane wave solutions of the Hamiltonian (\ref{TMD_Hamiltonian}) in a system without edge and show that for energies around valence band extrema they are expressed by the formula (\ref{plane_wave}). For simplicity we consider only spin up electrons in the $K$ valley (i.e. $\tau =1$) and suppress the spin and valley indexes below in this section. Therefore, components $\psi_{1,2}$ of a plane wave solution \mbox{$\Psi_{\bf p} = \left[e^{i{\bf p}{\bf r}/\hbar}/(L_xL_y)^{1/2}\right](\psi_1,\psi_2)^{\rm T}$} satisfy the system:
\begin{equation}
\left\{
\begin{split}
\left(m+\Delta_c - \varepsilon\right)\psi_1 + v(\tau p_x - ip_y)\psi_2 = 0\\
v(\tau p_x + ip_y)\psi_1 + \left(-m+\Delta_v - \varepsilon\right)\psi_2 = 0.
\end{split}
	\right.
\end{equation}
Together with normalization condition 
\begin{equation}
\int_{-L_x/2}^{L_x/2}dx\int_{-L_y/2}^{L_y/2}dy\Psi^*_{\bf p}\Psi_{\bf p} = 1, \nonumber
\end{equation}
solutions for the amplitudes can be represented as follows:
\begin{equation}\label{AppAmplitudes}
\begin{split}
\psi_1 = \frac{\left(\varepsilon_{c,v} + m -\Delta_v\right)}{\left[\left(\varepsilon_{c,v} + m -\Delta_v\right)^2 + (vp)^2\right]^{1/2}}, \\
\psi_2 = \frac{vpe^{i\theta_{p}}}{\left[\left(\varepsilon_{c,v} + m -\Delta_v\right)^2 + (vp)^2\right]^{1/2}}, \\ 
\end{split}
\end{equation}
where energies for the plane wave solutions read as follows:
\begin{equation}\label{App_spectra}
\varepsilon_{c,v} = \left(\Delta_v+\Delta_c\right)/2 \pm \sqrt{\widetilde{m}^2 + (vp)^2}, 
\end{equation} 
here sign $+(-)$ before square root corresponds to bulk states in conduction (valence) band respectively. Finally, expanding energies of valence band states (\ref{App_spectra}) in expressions for the amplitudes (\ref{AppAmplitudes}) around the band extremum (i.e. at \mbox{$vp\ll 2\widetilde{m}$}): \mbox{$\varepsilon_v\approx -m +\Delta_v + (vp)^2/2\widetilde{m}$}, we obtain the following expression for the plane wave solutions in vicinity of valence band maximum: %\mbox{$\psi_1\approx -vp/2^{3/2}\widetilde{m}$}, \mbox{$\psi_2\approx e^{i\theta_{p}}/2^{1/2}$}, which are identical to that was used in the main body of the paper (see Eq.(\ref{plane_wave})).
\begin{equation}
\Psi_{\bf p}=\frac{1}{\sqrt{L_xL_y}}
\left(
\begin{array}{c}
-\dfrac{vp}{2\widetilde{m}} \\
\\
e^{i\vartheta_p}
\end{array}
\right)e^{i{\bf p}{\bf r}/\hbar},
\end{equation}
which is identical to Eq.(\ref{plane_wave}) used in the main text. For the states around the conduction band minimum \mbox{$\varepsilon_c\approx m + \Delta_c + (vp)^2/2\widetilde{m}$}, one can obtain the following expressions:
\begin{equation}
\Psi_{\bf p}=\frac{1}{\sqrt{L_xL_y}}
\left(
\begin{array}{c}
 1 \\
\\
\dfrac{vp}{2\widetilde{m}}e^{i\vartheta_p}
\end{array}
\right)e^{i{\bf p}{\bf r}/\hbar},
\end{equation} 

\section{Derivation of kinetic equation (\ref{kin_eq})}\label{AppA}

In order to derive the kinetic equation (\ref{kin_eq}) we first write down the Hamiltonian of the system under consideration in terms of second-quantization operators:
\begin{multline}\label{AppA_H0}
H_0 = \sum_{p_e}\varepsilon_e a^{+}_{p_e}a_{p_e} + \sum_{p_x,p_y}\varepsilon_{v,p_x,p_y}a^{+}_{v,p_x,p_y}a_{v,p_x,p_y} + \\ \sum_{p_x,p_y}\varepsilon_{c,p_x,p_y}a^{+}_{c,p_x,p_y}a_{c,p_x,p_y} + \sum_{{\bf q},\lambda=\pm}\hbar\omega\left(c^+_{{\bf q},\lambda}c_{{\bf q},\lambda} + \frac{1}{2}\right),
\end{multline}
where $a_{p_e} (a^{+}_{p_e})$ is annihilation (creation) operator of the edge state (\ref{ES_function}) with energy $\varepsilon_{e}$, $a_{v/c, p_x, p_y}$ $(a^+_{v/c, p_x, p_y})$ is annihilation (creation) operator of the valence/conduction band state (\ref{bulk_solution}) with energy $\varepsilon_{v/c,p_x,p_y}$,   $c_{{\bf q},\lambda} (c^+_{{\bf q},\lambda})$ is annihilation (creation) operator of photon with clockwise ($\lambda=+$) or counter-clockwise ($\lambda=-$) polarisation and energy $\omega=cq$. In previous equation we suppress spin and valley indexes for brevity. In terms of the creation and annihilation operators of photon field vector-potential reads as follows:
\begin{equation}
{\bf A}({\bf r}, t) = \sum_{{\bf q},\lambda=\pm}\sqrt{\frac{2\pi c^2\hbar}{n_r^2\omega\Omega}}\left[c_{{\bf q},\lambda}{\bf e}_{\lambda}e^{i({\bf q r} - \omega t)} + c^{+}_{{\bf q},\lambda}{\bf e}^*_{\lambda}e^{-i({\bf q r} - \omega t)}\right],
\end{equation}  
where $\Omega$ is quantization volume, polarisation unit vectors ${\bf e}_{{\bf q},\pm} = [1/\sqrt{2}]\left(\cos\alpha_q \pm i\sin\alpha_q\cos\theta_q, -\sin\alpha_q \pm i\cos\alpha_q\cos\theta_q,\mp i\sin\theta_q\right)$ (spherical angles $\alpha_q$, $\theta_q$ characterize direction of the photon wave-vector ${\bf q} = q\left(\sin\alpha_q\sin\theta_q,\cos\alpha_q\sin\theta_q,\cos\theta_q\right)$). Interaction of electrons with electromagnetic field reads as follows:%\left({\bf e}_1 \pm i{\bf e}_2 \right)
\begin{equation}\label{App_interaction}
V_{int} = \sum_{p_e,p_x,p_y,{\bf q},\lambda=\pm} \left[V^{(\lambda,\tau)}_{\varepsilon_v,\varepsilon_e}({\bf q})a^{+}_{p_e}c_{{\bf q},\lambda}a_{v,p_x,p_y}+ h.c.\right],
\end{equation}
where we take into account only transitions between valence band electrons and edge states that are determined in dipole approximation by the matrix elements:
\begin{equation}
\begin{split}\label{App_field_matrix_element}
V^{(\lambda,\tau)}_{\varepsilon_v,\varepsilon_e}({\bf q}) = \tau\delta_{p_e,p_x}\frac{ve}{c}\sqrt{\frac{2\pi c^2\hbar}{n_r^2\omega\Omega}}\frac{C_e\sqrt{L_x}}{\sqrt{2L_y}}\frac{2ik_y(\varepsilon,p_e)}{\kappa^2 + k_y^2(\varepsilon,p_e)}\times\qquad\qquad \\
\left[\frac{e^{-i\tau\alpha_q}}{\sqrt{2}}\left(1+\tau\lambda\cos\theta_q\right) + \frac{e^{i\tau\alpha_q}}{\sqrt{2}}\left(1-\tau\lambda\cos\theta_q\right)\frac{\hbar v\kappa + \tau v p_e}{2\widetilde{m}a}\right],\quad
\end{split}
\end{equation}
From the previous formula it follows that ratio of probabilities for induced transitions (at normal incidence of light $\cos\theta_q=1$) in the two valleys for definite polarisation has an order of $\left(\hbar v\kappa/2\widetilde{m}a\right)^2\approx 1/4\ll 1$ at the boundary parameter values $|a|\approx 1$. 
 
Now we introduce Keldysh Green functions of electrons $G^{\alpha\beta}_{\nu}(t,t')=-i\left\langle T_{C}\left\{a_{\nu}(t^{\alpha})a_{\nu}^{+}(t'^{\beta})\right\} \right\rangle$ ($\nu$ means quantum numbers of edge or bulk state) and photons $D^{\alpha\beta}_{\bf q,\lambda }(t,t')=-i\left\langle T_{C}\left\{c_{\bf q,\lambda }(t^{\alpha})c_{\bf q,\lambda }^{+}(t'^{\beta})\right\} \right\rangle$ ($\alpha,\beta=\pm$). Following standard procedure\cite{bib:Arseev} we obtain a kinetic equation  for Green function that determines population distribution of edge state ($G^{<}_{e}(t,t)=G^{-+}_{e}(t,t)$):
\begin{multline}\label{App_gen_kin_eq}
i\frac{\partial }{\partial t}G^{<}_e(t,t) =  \\
\int_{-\infty}^{+\infty}\Sigma^R_{ee}(t,t_1)G^{<}_e(t_1,t)dt_1 + \int_{-\infty}^{+\infty}\Sigma^<_{ee}(t,t_1)G^{A}_e(t_1,t)dt_1 - \\
\int_{-\infty}^{+\infty}G^{R}_e(t,t_1)\Sigma^<_{ee}(t_1,t)dt_1 -\int_{-\infty}^{+\infty}G^{<}_e(t,t_1)\Sigma^A_{ee}(t_1,t),
\end{multline}
where $G^R=G^{--} - G^{-+}$, $G^A=G^{--} - G^{+-}$. In the Eq. (\ref{App_gen_kin_eq}) we calculate self-energies in the second order in perturbation (\ref{App_interaction}): 
\begin{multline}\label{App_self_energy}
\Sigma^{R}_{ee}(t_1,t_2)=\Sigma^{--}_{ee}(t_1,t_2) + \Sigma^{-+}_{ee}(t_1,t_2) = \\ 
i\sum_{ {\bf p},{\bf q},\lambda}\left|V^{(\lambda,\tau)}_{\varepsilon_v,\varepsilon_e}({\bf q})\right|^2 \left[ D^{--}_{0_{{\bf q},\lambda}}(t_1,t_2)G^{--}_{0_{{\bf p},\varepsilon_v}}(t_1,t_2) - \right. \\ 
\left. D^{-+}_{0_{{\bf q},\lambda}}(t_1,t_2)G^{-+}_{0_{{\bf p},\varepsilon_v}}(t_1,t_2) \right] =  \\
-i\theta\left(t_1 - t_2\right) \times  \\
\sum_{{\bf p},{\bf q},\lambda} \left|V^{(\lambda,\tau)}_{\varepsilon_v,\varepsilon_e}({\bf q})\right|^2\left[n_0\delta_{\lambda,+}\delta_{{\bf q},{\bf q}_0} + \left(1 - f_v\right) \right]e^{-i(\varepsilon_v + \omega)(t_1-t_2)}, \\
\Sigma^{<}_{ee}(t_1,t_2) = -\Sigma^{-+}_{ee}(t_1,t_2) =  \\ 
i\sum_{{\bf p}} \left|V^{(+,\tau)}_{\varepsilon_v,\varepsilon_e}({\bf q}_0)\right|^2 n_0f_{v}e^{-i(\varepsilon_v + \omega_0)(t_1-t_2)}, \\
\Sigma^{R}_{ee}(t_1,t_2) = \left[\Sigma^{A}_{ee}(t_2,t_1)\right]^*. 
\end{multline}
where the terms with wave vector ${\bf q}_0=(0,0,-q_0)$ describe transitions induced by illuminated light with frequency $\omega_0=cq_0$ and clockwise polarisation, $n_0=-iD^{-+}_{0_{{\bf q}_0,+}}(t,t)$ is the number of the illuminated light quanta, the term proportional to $1-f_v$ (where $f_v=-iG^{-+}_{0_{{\bf p},\varepsilon_v}}(t,t)$ is Fermi-Dirac distribution function in valence band) concerns with spontaneous recombination processes. After substitution of Green functions of zero approximation \mbox{$G^A_e(t_1,t_2)=\left[G^{R}_e(t_2,t_1)\right]^*=i\theta(t_2-t_1)e^{i\varepsilon_e(t_2-t_1)}$}, $G^{<}_{e}(t_1,t_2)=if_e(t) e^{i\varepsilon_e(t_2-t_1)}$ in Eq. (\ref{App_gen_kin_eq}) and accounting for only contributions from poles at integration over time we arrive to a kinetic equation similar to that used in the main text (\ref{kin_eq}):
\begin{widetext}
\begin{equation}\label{App_kin_eq}
\frac{\partial f_e}{\partial t} = \frac{I\Omega}{c\hbar\omega_0}\frac{2\pi}{\hbar}\sum_{\bf p} \left|V^{(+,\tau)}_{\varepsilon_v,\varepsilon_e}({\bf q}_0)\right|^2\left[f_v - f_e\right]\delta\left(\varepsilon_v - \varepsilon_e+\hbar\omega_0\right) -  f_e \frac{2\pi}{\hbar}\sum_{{\bf p}, {\bf q},\lambda=\pm}\left|V^{(\lambda,\tau)}_{\varepsilon_v,\varepsilon_e}({\bf q})\right|^2\left[1 - f_v\right]\delta\left(\varepsilon_v - \varepsilon_e + \hbar\omega\right) - \frac{f_e - f_{eq}}{\tau_R},
\end{equation}
\end{widetext}
where we add the term with phenomenological relaxation time $\tau_R$ as in the main text, $n_0 = I\Omega/c\hbar\omega_0$.

\section{General relation for quasi Fermi-levels}

In the section we will obtain a general relation for quasi Fermi-levels of electrons in valence and edge bands under illumination of the light. Introducing the density of valence band states with definite momentum (\ref{DOS_v}) we rewrite Eq. (\ref{App_kin_eq}) in the following form:
\begin{widetext}
\begin{align}
\frac{I}{I_0}\left[\delta_{\tau,1} + \delta_{\tau,-1}\left(\frac{\hbar v \kappa_e+\tau vp_e}{2\widetilde{m}a}\right)^2\right]\left[f_{v}(\varepsilon-\hbar\omega_0) - f_e(\varepsilon)\right]\frac{\sqrt{\varepsilon_{t} - \frac{v^2p_e^2}{2\widetilde{m}} - \varepsilon + \hbar\omega_0}}{\left[\frac{(\hbar v \kappa_e)^2}{2\widetilde{m}} + \varepsilon_{t} - \frac{v^2p_e^2}{2\widetilde{m}} - \varepsilon + \hbar\omega_0\right]^2} - \nonumber \\
\frac{f_e(\varepsilon)}{12\pi^2(\hbar\omega_0)^2}\left[1 + \left(\frac{\hbar v \kappa_e+\tau vp_e}{2\widetilde{m}a}\right)^2 \right] \int_0^{+\infty} d(\hbar\omega) \frac{\hbar\omega\left[1 - f_v(\varepsilon - \hbar\omega)\right]\sqrt{\varepsilon_{t} - \frac{v^2p_e^2}{2\widetilde{m}} - \varepsilon + \hbar\omega}}{\left[\frac{(\hbar v \kappa_e)^2}{2\widetilde{m}} + \varepsilon_{t} - \frac{v^2p_e^2}{2\widetilde{m}} - \varepsilon + \hbar\omega\right]^2} - \nonumber \\
\frac{(1+a^2)n_r^2c^3\sqrt{2\widetilde{m}}}{4a^2\pi\tau_R(ve)^2\omega_0^2\hbar v\kappa_e }\left[f_e(\varepsilon) - f_{eq}(\varepsilon)\right]=0
\end{align}
\end{widetext}
To proceed further analytically, we consider low temperature limit ($T\ll \min\left (\left|\mu_{e,\tau}-\varepsilon_F\right|,\left|\mu_{v,\tau}-\varepsilon_t\right| \right)$ to treat Fermi-functions as step-like ones. Then we integrate the above equation over energy and arrive a final relation between quasi Fermi-levels $\mu_{e,\tau}, \mu_{v,\tau}$:
\begin{widetext}
\begin{align}\label{App_relation}
\frac{I}{I_0}\left[ \delta_{\tau,1} + \delta_{\tau,-1}\left( \frac{\hbar v \kappa_F}{2\widetilde{m}a} \right)^2 \right]
\left\{ 
\arctan \left[ \frac{\kappa_F \left( k_{y0}(\mu_{e,\tau}-\hbar\omega_0) - k_{y0}(\mu_{v,\tau})\right )}{\kappa_F^2 + k_{y0}(\mu_{e,\tau}-\hbar\omega_0)k_{y0}(\mu_{v,\tau})}\right] + \frac{\kappa_F k_{y0}(\mu_{v,\tau})}{\kappa_F^2 + k^2_{y0}( \mu_{v,\tau})} \right\} - \nonumber \\
\theta(\mu_{e,\tau}-\varepsilon^{*})\frac{\kappa_F\left[1 + \left(\frac{\hbar v \kappa_0}{2\widetilde{m}a}\right)^2\right]}{12\pi^2\kappa_0\left(\hbar\omega_0\right)^2} \left\{ 6(\hbar v)^2\kappa_0 k_{y0}(\mu_{v,\tau})\frac{\mu_{e,\tau} - \varepsilon^{*}}{2\widetilde{m}} - \frac{\kappa_F k_{y0}(\mu_{v,\tau})}{\kappa_F^2 + k^2_{y0}( \mu_{v,\tau})} \left[\left(\mu_{e,\tau} - \mu_{v,\tau}\right)^2 - \left(\varepsilon^{*} - \mu_{v,\tau}\right)^2\right] + \right. \nonumber \\ \left. \arctan\left(\frac{k_{y0}(\mu_{v,\tau})}{\kappa_0}\right) \left[\left(\mu_{e,\tau} - \varepsilon_{t} - 3\frac{(\hbar v \kappa_0)^2}{2\widetilde{m}} \right)^2 - \left(\varepsilon^* - \varepsilon_{t} - 3\frac{(\hbar v \kappa_0)^2}{2\widetilde{m}} \right)^2\right]\right\}
- %\nonumber \\
\frac{(1+a^2)n_r^2c^3}{4a^2\pi\tau_R\omega_0^2 (ve)^2  }\left[\mu_{e,\tau} -\varepsilon_F\right]=0,
\end{align}
\end{widetext}
where $k_{y0}(\varepsilon) = k_y(\varepsilon,0)$, $\kappa_F = \kappa(\varepsilon_F)$, $\kappa_0 = \kappa(\varepsilon_e(0))$, and $\varepsilon^{*}$ is a minimal energy of electrons filling ESs that can radiatively recombine with holes in the valence band. At integration we neglected by dependence of ES decay length and terms proportional to $vp_e/2\widetilde{m}$ on the energy. In the limit $\rho_e/\rho_v\ll 1$ we can neglect by the terms proportional to $k_{y0}(\mu_{v,\tau})$ in Eq.(\ref{App_relation}), which leads us to the Eq.(\ref{equation_mu_el}) used in the main text of the manuscript.

\bibliography{Bibl}

\end{document}